\documentclass[twocolumn,showpacs,floatfix,superscriptaddress,citeautoscript,cite]{revtex4}
\usepackage{graphicx}
\usepackage{xfrac}

\begin{document}

	\author{J. Sivek}
		\email{jozef.sivek@ua.ac.be}
	\author{O. Leenaerts}
		\email{ortwin.leenaerts@ua.ac.be}
	\author{B. Partoens}
		\email{bart.partoens@ua.ac.be}
	\author{F. M. Peeters}
		\email{francois.peeters@ua.ac.be}

	\affiliation{Departement Fysica, Universiteit Antwerpen, Groenenborgerlaan 171, B-2020 Antwerpen, Belgium}

	\title{Adsorption of titanium and titanium dioxide on graphene: n and p-type doping}

	\date{\today}


\begin{abstract}
\textit{Ab initio} calculations within the density-functional theory formalism are performed to investigate the ground state, electric charge doping, and electronic properties of titanium and titanium dioxide monolayers adsorbed on a graphene surface. A new ground state structrure of Ti monolayer adsorbed on graphene is reported which is shown to be stable up to T = 500 K. Effects due to lower and higher Ti adatoms coverage are studied. We find that the adsorbed Ti provides a strong n-type doping which supports recent experimental observations. On the other hand, TiO$_{2}$ can induce both p- and n-type doping in the carbon monolayer depending on whether oxygen or titanium atoms are closest to the substrate. We identify the structures which are responsible for the experimentally observed autocompensation mechanism that leads to the reversion of adsorbate effects after oxidation of the adsorbed Ti.

\end{abstract}

\maketitle

\section{Introduction}

The interaction of graphene with metal adsorbates
is one of the important research topics since the first
successful experimental exploration of graphene.\ \cite{novoselov_2004}
Virtually any device assembly incorporating graphene will
necessarily include graphene-metal contacts.

Titanium has been widely used for metal contacts on
graphene with good adsorption properties.\ \cite{robinson_2011}
Moreover, the interaction between titanium 
and graphitic surfaces and nanotubes
has attracted a lot of attention.\ \cite{dag_2004}
Ti atoms are able to act as adsorption centers for
molecular binding.\ \cite{carrillo_2009}
Furthermore, Ti coated nanotubes and graphene
have also been proposed for high-capacity
hydrogen storage solution.\ \cite{yildirim_2005, liu_2010}

However, a recent DFT study,\ \cite{rojas_2007}
as well as a very recent experimental observation,\ \cite{mccreary_2011}
showed strong structural changes of the adsorbed Ti layer
on graphene due to the interaction with oxygen.
Experimentally, it was shown that the adsorbed titanium
on graphene leads to substantial
n-type doping and a reduction of graphene's 
mobility. The subsequent exposure of the samples to oxygen
has restored graphene's gate dependent conductivity 
to almost intrinsic values, effectively cancelling
any previous metal--graphene interaction.\ \cite{mccreary_2011}

Motivated by these experimental observations we perform
\textit{ab initio} calculations to investigate titanium
and titanium dioxide monolayers adsorbed on graphene.
For the identified ground state structures, the electronic
band structure is calculated and a charge population analysis is performed
to investigate the character of the charge carrier doping in graphene.
We find a strong n-type doping of graphene induced
by adsorbed Ti and identify the highest possible monolayer
coverage with stoichiometry Ti$_{3}$C$_{8}$ .
When TiO$_{2}$ monolayer crystals are aligned on
top of the graphene surface we observe
p- or n-type doping of graphene,
depending on the nature of the
atoms that are exposed closest to graphene with oxygen acting as an acceptor
and titanium as a donor.

This paper is organized as follows.
First, we describe the computational details of our {\it ab initio}
calculations, followed by an investigation of
the ground state structure of a titanium monolayer on graphene.
Next the character of the charge transfer between
adsorbed Ti and the graphene substrate is examined.
Further model titanium dioxide monolayer
structures are investigated.
To conclude, we report titanium dioxide structure
dependent doping of graphene 
and link this observation with the experimentally observed recovery of
the graphene gate dependent conductivity
after the oxidation of the adsorbed Ti atoms.

\section{Calculations}

All our calculations were performed within the density functional theory (DFT)
formalism as implemented in the VASP package with usage of the local spin
density approximation (LSDA) for the exchange-correlation functional.
We made use of the projector augmented wave method \cite{bloch_1994-paw}
and a plane-wave basis set with an energy cutoff of 500 eV.
The complete relaxation of atomic positions and supercell size
was performed up to the level that the forces are smaller than 0.01~eV~\AA$^{-1}$.

For all the used supercells, a sampling of the Brillouin zone was done
with the equivalent of a $28\times28\times1$ Monkhorst--Pack \cite{monkhorst_1976}
$k$-point grid for the graphene unit cell (containing two carbon atoms).
Spin polarization was included in the calculations.

Because periodic boundary conditions were applied in all three dimensions,
the height of the supercell was set to 20~{\AA}, to include enough vacuum
to minimize the interaction between adjacent layers, and
dipole corrections were used.
All reported quantitative results of the charge transfer
were calculated with the iterative Hirshfeld
charge population analysis.\ \cite{bultinck_2007-ih}

\section{Results and Discussion}

The binding energies are calculated with respect
to intrisic monolayer graphene and the isolated atom/molecule
of which the binding energy is reported. The formation
energy for titanium covered graphene
is calculated in relation to intrinsic monolayer graphene
and the energy the of Ti atoms in a free standing
monolayer titanium crystal with hexagonal symmetry.

\subsection{Properties of titanium monolayer on graphene}

Properties of single Ti atom adsorption on a graphene
surface were already investigated and the most preferable
adsorption site for a Ti atom was found to be the hollow site with
a binding energy of -1.58 eV per Ti atom.
The top and bridge adsorption site were found to be energetically less favorable by
0.62 eV 
and result in metastable states.\ \cite{sevincli_2008-tm_on_g}

\begin{figure}[h]
  \centering
\includegraphics[width= 8.5 cm]{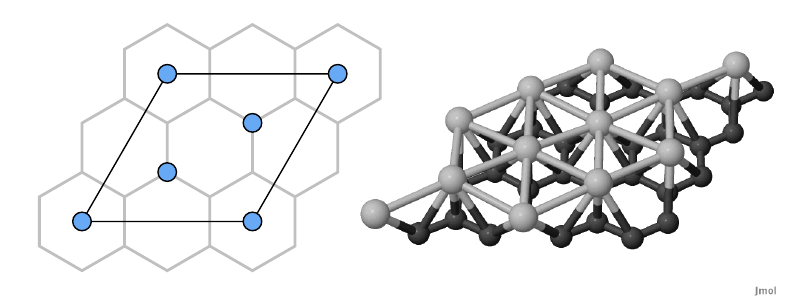}
	\caption{\label{fig_ti_g_structures}
	Structure g2x2-3Ti: Ti monolayer on a graphene sheet,
	3 Ti atoms (light) per 8 C atoms (dark, displayed
	as a mesh in schematic picture on the left).
	}
\end{figure}

Also high density coverage of graphene with titanium
has been previously investigated and some ground state
structures were proposed which have
stoichiometry TiC$_{8}$ and TiC$_{2}$ with Ti atoms
placed on hollow sites creating monolayer metalic
coverage.\ \cite{rangel_2009-h2o_on_ti-g, zanella_2008-elmag_of_ti_fe_on_g}
However, we report here a different ground state structure.
As written above, a single Ti atom placed on top of a free
standing graphene surface is strongly bonded at the hollow site
with a diffusion barrier high enough to prevent its
motion on the surface even at room temperature.
This fact suggests that the most preferable
high coverage structure for titanium on graphene
will be the one in which above every single hollow site
there is one Ti atom (i.e., one Ti atom per two C atoms).
Even the lattice constants compare favourably
for this structure with a mismatch of only 5~\%~(see~Table~\ref{table_pristine_structures}).

\begin{figure}[h]
  \centering
\includegraphics[width= 8.5 cm]{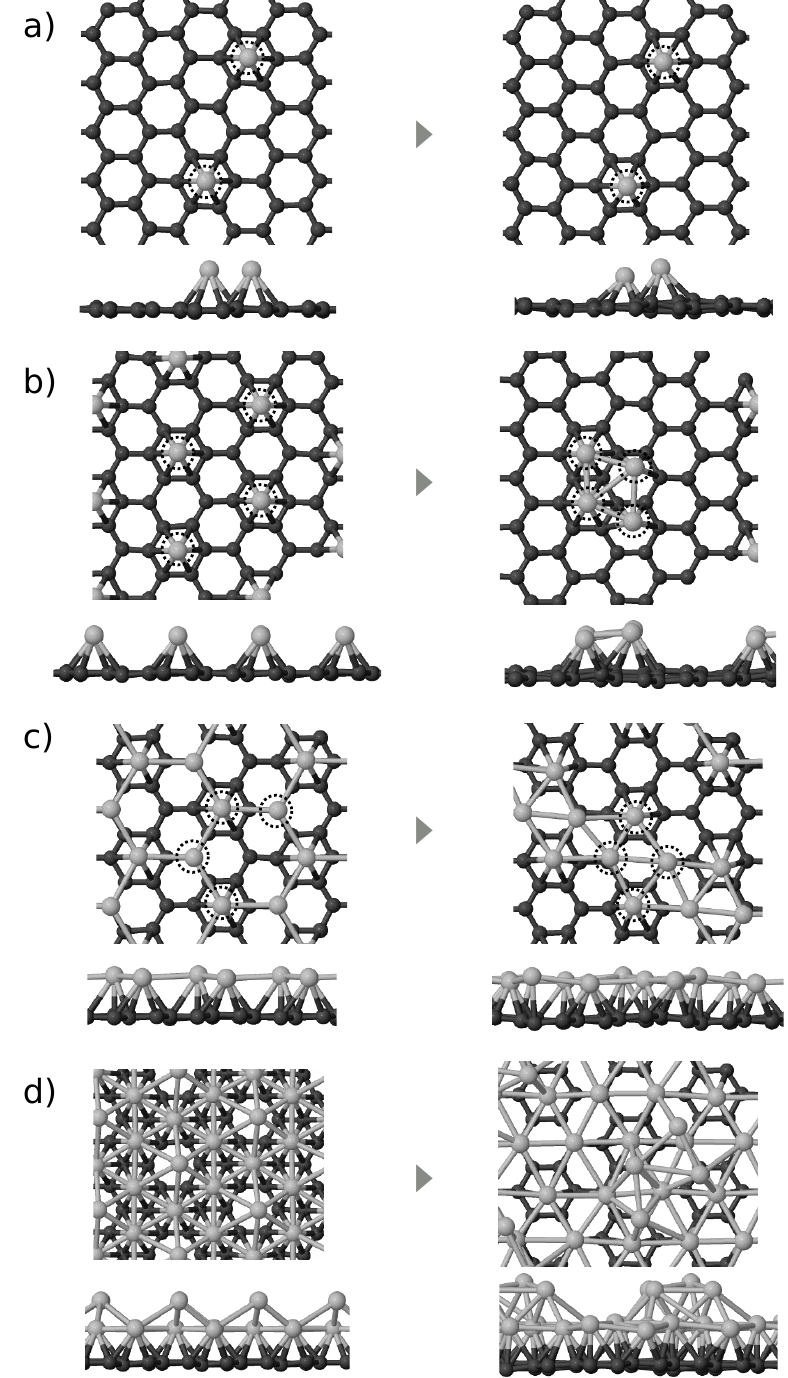}
	\caption{\label{fig_ti_g_md}
	Top and side view of different graphene structures with
	different amount of titanium coverage before (left)
	and after (right) 1 ps of relaxation
	at $T = 500$ K. The systems consist of
	(a) 2, (b) 4, (c) 8 and (d) 16 Ti adatoms (light symbols)
	placed on the graphene sheet in an unit
	cell with 32 C atoms (dark symbols).
	}
\end{figure}

We have calculated binding energies
for a variety of structures with stoichiometry TiC$_{2}$
and we also investigated the structure g2x2-3Ti
(displayed in Fig. \ref{fig_ti_g_structures})
with a lower amount of Ti atoms per C atom (three Ti atoms per eight C atoms)
and a larger lattice mismatch of -9 \% (Ti atoms are spaced further
apart from each other in comparison to the titanium monolayer crystal).
The calculated formation energy of -0.93 eV per Ti for g2x2-3Ti system
was found to be lowest among all the investigated structures.
The lowest formation energy for the TiC$_{2}$ system
was found to be -0.76 eV per Ti, for the
structure consisting of adatoms
placed above hollow sites with every
fourth Ti atom displaced further
from the graphene substrate (see Fig. \ref{fig_ti_g_md}(d) initial state).
Substantial formation energy advantage makes the g2x2-3Ti
system virtually the only plausible structure for titanium
atoms on graphene at large coverage densities.

We have been surprised by this finding that
the structure g2x2-3Ti has the lowest binding energy
among studied cases. This has motivated us to perform
DFT based molecular dynamics simulations of systems with different
Ti coverage densities to identify the ground state
configurations that may have been overlooked and/or
to support the aforementioned findings.

The simulated systems consist of a 4x4 graphene supercell
with 2, 4, 8, and 16 Ti adatoms, as displayed
in Figs.~\ref{fig_ti_g_md}(a-d).
At the temperature of 500 K the different systems were allowed to evolve
to more favourable states within a time window exceeding 1 ps.

The system with the highest concentration coverage (TiC$_{2}$),
with Ti atoms initially placed above every hollow site,
does not remain a monolayer crystal.
However, something interesting happens: the Ti atoms 
rearrange themselves so that they create one monolayer
on the graphene surface with the same structure as g2x2-3Ti
while the redundant Ti atoms create
a second layer.

The systems with lower concentration of Ti atoms per C atoms
(TiC$_{8}$ and TiC$_{4}$, Figs. \ref{fig_ti_g_md}(b) and
\ref{fig_ti_g_md}(c), respectively)
have also undergone structural changes.
The separation of adsorbed Ti atoms was
too small to prevent atoms from mutual interaction.
While adatoms remained adsorbed as a monolayer
we have observed the creation of clusters and local
rearangements. Those structural changes occured despite
the presence of the diffusion barrier for adatoms adsorbed
on the hollow site (clearly visible for TiC$_{8}$, Fig. \ref{fig_ti_g_md}(b)).
Again the comparison with the structure g2x2-3Ti
reveals remarkable similarities in local arrangements
of Ti adatoms, which is more pronounced in the 
structure TiC$_{4}$ (Fig. \ref{fig_ti_g_md}(c))
with higher adatom coverage.

The last of the systems investigated using MD calculations
is the one with the lowest Ti concentration  - TiC$_{16}$ (Fig. \ref{fig_ti_g_md}(a)).
We found that this case is the only one
that maintained its structural
properties during the MD simulations.
As can be seen in Fig. \ref{fig_ti_g_md}(a)
the adsorbed atoms are separated by
two carbon hexagonal rings.
The separation distance of 6.5~\AA~is found
to be sufficient to suppress
the mutual adatom interaction observed
in the previous cases. The Ti atoms
were trapped at the hollow sites and do not
pass across the diffusion barrier.

The aforementioned results of the MD calculations can be
summarized as follows:
(i) Ti atoms placed on a graphene surface can
overcome diffusion barriers (for the hollow site)
by mutual close range interaction (up to $\pm 6.5 \mathrm{\AA}$),
(ii) the maximal titanium monolayer coverage of graphene
has stoichiometry Ti$_{3}$C$_{8}$ and corresponds
to the strucure g2x2-3Ti (Fig. \ref{fig_ti_g_structures}).

Experimentally \cite{mccreary_2011} a strong
charge transfer from the Ti layer
to graphene was observed.
Charge population analysis performed on the investigated
systems has revealed the same trend of strong to moderate
charge transfer from the Ti atoms to the underlying
graphene for low and high concentrations, respectively.
However, the quantitative and even qualitative values for charge transfer
were found to be inconsistent among the different methods we used
(including the iterative Hirshfeld charge population analysis, 
modified iterative Hirshfeld method and the Bader and Voronoi cells
charge population analysis).\ \cite{bultinck_2007-ih, leenaerts_2008-charge, bader_1991, tang_2009, celia_2004-voronoi}

\begin{figure}[h]
  \centering
\includegraphics[width= 8.5 cm]{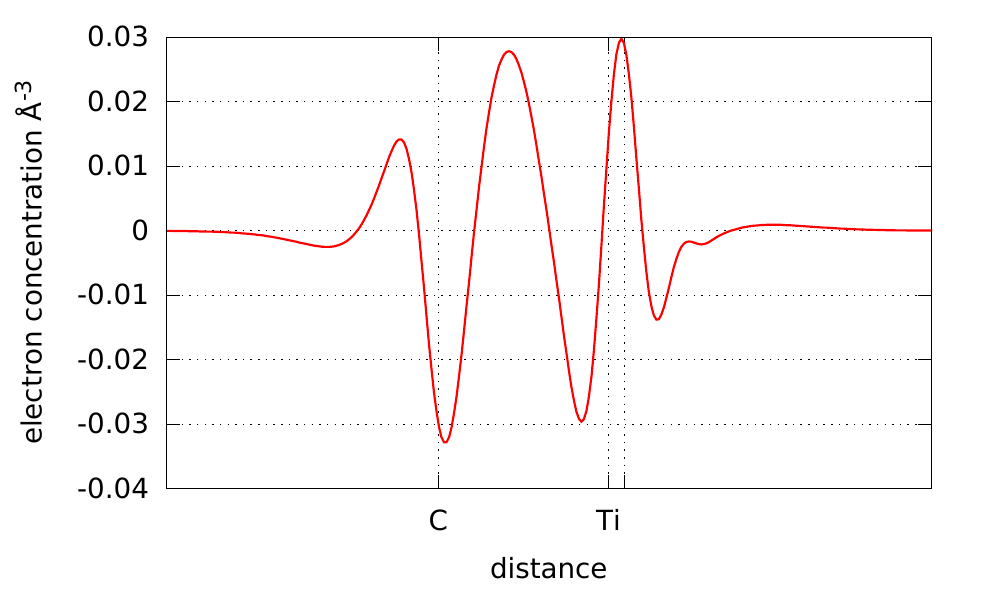}
	\caption{\label{fig_ti_g_def_charge}
	Plane-averaged deformation electron density of g2x2-3Ti configuration
	obtained from the difference between the electron density of
	the g2x2-3Ti structure and the electron densities
	of the subsystems consisting of only carbon or titanium	atoms.
	}
\end{figure}

An alternative to obtain directly the
charge transfer is by calculating
the dipole moment after Ti adsorption.
However, the value of the electric dipole
moment for the g2x2-3Ti structure is 0.087 e\AA,
suggesting a small charge transfer from graphene to Ti.
This corresponds neither to the experimental
observation of n-type doping, nor with
the charge transfer analyses discussed
above. As shown by the plane-averaged deformation
electron density in Fig. \ref{fig_ti_g_def_charge},
the strong oscillations in the charge density
do not clearly indicate actual charge transfer
nor a dipole that will identify such transfer.
The explanation for opposite total electric dipole
moment lies in the dominant polarisation
of the carbon layer induced by the positively 
charged chemisorbed Ti monolayer. The surplus of charge
located on the C atoms is shifted towards
the Ti atoms thus creating a dipole moment
counteracting and exceeding the dipole moment
based on the negative charge located on the carbon layer
and the positively charged Ti atoms.
We expect that the polarisation of the carbon layer
to be the probable reason for the inconsistent
results obtained by the aforementioned
charge population analysis methods.

\begin{figure}[h]
  \centering
\includegraphics[width= 8.5 cm]{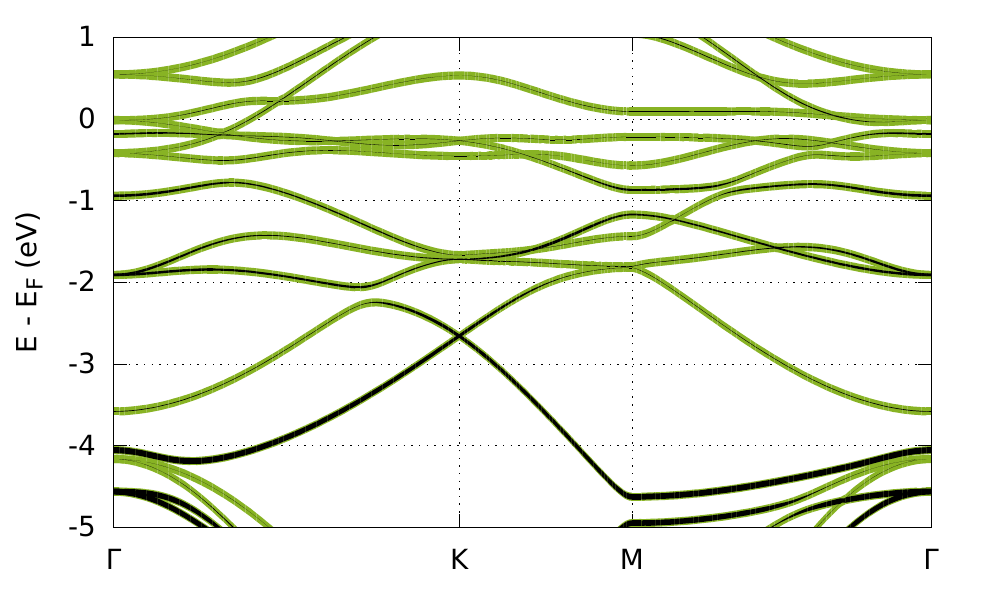}
	\caption{\label{fig_ti_g_bands}
	Band structure of the g2x2-3Ti configuration. 
	The energies are relative to the Fermi level (E$_F$ = 0).
	The Dirac point lies -2.61 eV below the Fermi level showing strong n-type doping.
	The amount of carbon p$_z$ character is indicated by the blackness of the bands.
	}
\end{figure}

\begin{table}[h]

	\caption{
	Lattice constants and workfunctions.
	\label{table_pristine_structures}}

\begin{tabular}{ l c c }
	\hline \hline
	structure                   & $\Phi$ (eV) & a (\AA)  \\
	\hline
	graphene                    & 4.49 & 2.447 \\
	hexagonal monolayer Ti      & 4.72 & 2.566 \\
	rectangular monolayer TiO2  & 8.54 & 3.141 \\
	hexagonal monolayer TiO2    & 8.51 & 2.945 \\
	\hline \hline
\end{tabular}
\end{table}

Since the direct charge transfer calculations proved to be
insufficient in determinig the type of doping we also performed
electronic band structure calculations and calculated the workfunction
of the pristine and hybrid materials.
As can be seen from the electronic
band structure of the g2x2-3Ti configuration (see Fig. \ref{fig_ti_g_bands}),
strong n-type doping occurs to graphene. The Dirac
cone, still present in the electronic band dispersion,
is shifted by -2.61 eV below the Fermi level.
The calculation of the workfunction reveals the same type of doping,
albeit in a partially counterintuitive way.
As shown in Table \ref{table_pristine_structures} the work function
of free-standing graphene is 4.49 eV, while the work function 
of an isolated hexagonal titanium monolayer is 4.72 eV.
From a simple comparison of these values there is no
precondition for charge transfer to occur from Ti
to the graphene substrate. However, for closely separated
metal-graphene structures (2.1 \AA~for g2x2-3Ti) the 
chemical interaction prevails and is responsible for
n-type doping even when the workfunction of the metal is larger than
the graphene work function.\ \cite{giovannetti_2008}
In the final system the computed workfunctions on
graphene (3.97 eV) and titanium side (4.72 eV) indicate
a charge transfer to graphene. The workfunction difference
increases as compared to the pristine values (0.81 eV compared to 0.23 eV).

\subsection{Properties of Titanium dioxide monolayer on graphene}

\begin{figure}[h]
  \centering
\includegraphics[width= 8.5 cm]{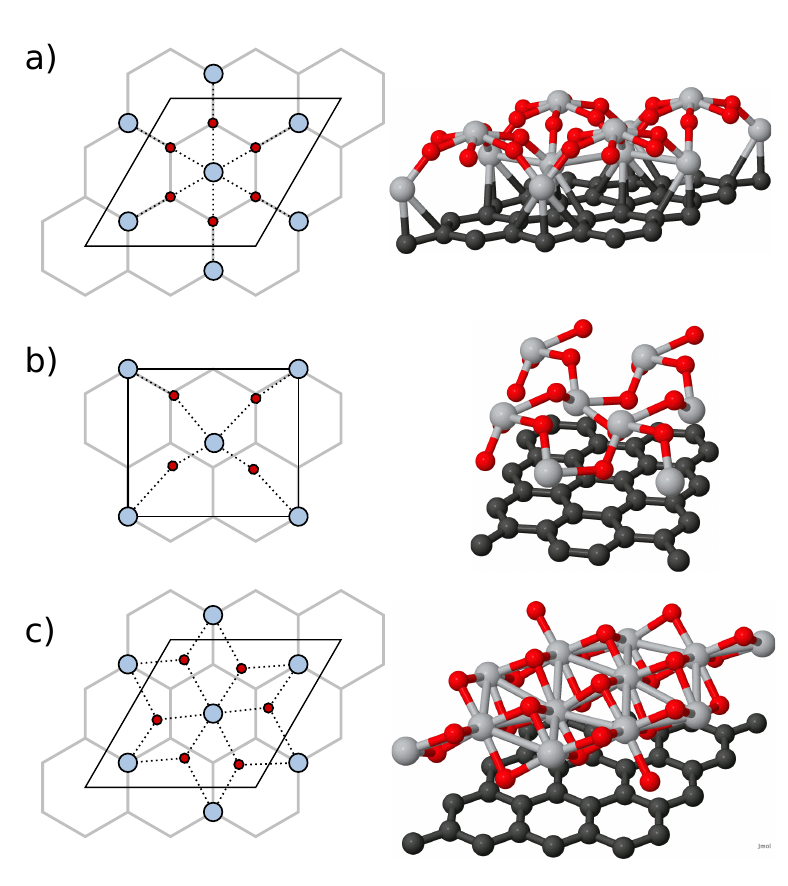}
	\caption{\label{fig_tio2_g_structures}
	Model structures for a TiO$_2$ monolayer on a graphene sheet.
	The systems (a) g2x2-aligned~hex-TiO$_{2}$ and (c) g2x2-hex-TiO$_{2}$
	contain 3 TiO$_2$ molecules per 8 C atoms,
	whereas (b) the g-sq8-TiO$_{2}$ configuration has 2 TiO$_2$ molecules per 8 C atoms.
	The carbon atoms are given by gray colour
	(a mesh in schematic picture on the left),
	Ti atoms are in light gray and O atoms are 
	shown in red (dark, small circles in schematic
	picture on the left).
	}
\end{figure}

Let us now consider the case of TiO$_{2}$ adsorption
on graphene. As a model approach for studying the interaction
of titanium dioxide on graphene surfaces we have chosen
monolayer TiO$_{2}$ films derived from anatase,
(101) with hexagonal and (001) with rectangular symmetry,
which were proven to be stable.\ \cite{evarestov_2010}
From the broad range of possible configurations
the structure g2x2-hex-TiO$_{2}$ with lattice mismatch
of about 4 \% (displayed in Fig. \ref{fig_tio2_g_structures}(c))
was found to be the most stable. Ti atoms
are aranged in a monolayer triangular lattice, similar
to the one found in the previously discussed
structure g2x2-3Ti, and oxygen atoms are placed
between every three neighbouring Ti atoms.
The TiO$_{2}$ monolayer is buckled due to
the alternating displacement of oxygen atoms
above or below the plane of Ti atoms.
The other displayed model configurations
in Figs. \ref{fig_tio2_g_structures}(a) and \ref{fig_tio2_g_structures}(b),
while being metastable, will be discussed below
as examples which lead to different
doping properties.

\begin{figure}[h]
  \centering
\includegraphics[width= 8.5 cm]{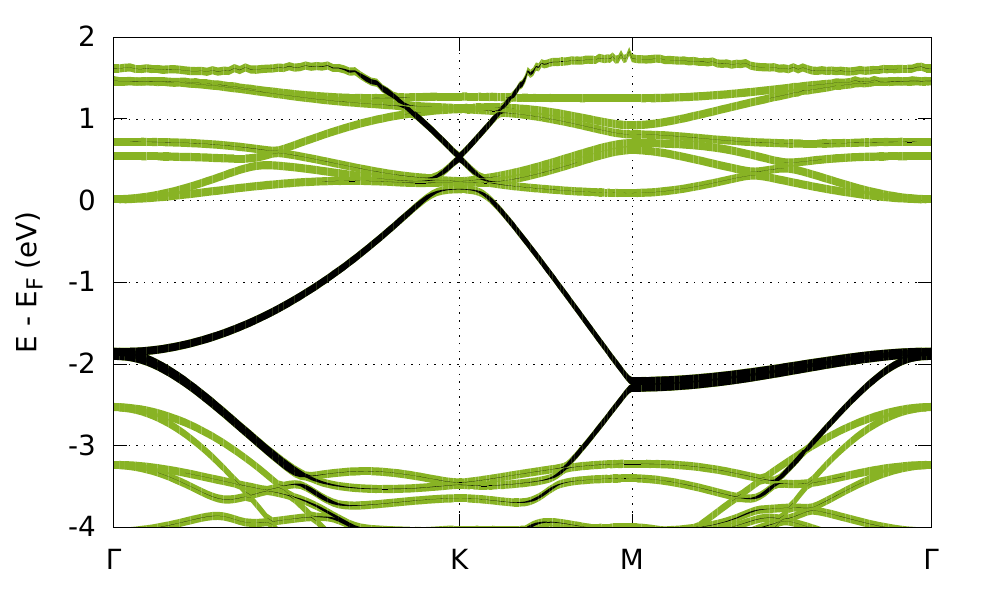}
	\caption{\label{fig_tio2_g_bands}
	Band structure of g2x2-hex-TiO$_{2}$ configuration. 
	The energies are relative to the Fermi level (E$_F$ = 0).
	The Dirac point lies 0.5 eV above the Fermi level showing p-type doping.
	The amount of carbon p$_z$ character is indicated by the blackness of the bands.
	}
\end{figure}

The charge population analysis performed on the
TiO$_{2}$-graphene structures were found to be much more
robust than for the titanium covered graphene.
The \mbox{Ti-O} bonds exhibit an ionic character with
Ti atoms gaining positive charge after loosing
two electrons in favour of the O atoms.
Besides the charge redistribution in the TiO$_{2}$
layer the g2x2-hex-TiO$_{2}$ structure shows
also p-type doping of the graphene layer.
The amount of charge transfered from graphene,
calculated with iterative Hirshfeld charge population analysis,
is 0.126 e per TiO$_{2}$ unit.

The character of the doping is also visible in 
the electronic band structure shown in Fig. \ref{fig_tio2_g_bands}
with a clear hole doping of graphene and a shift of
the Dirac cone above the Fermi level by 0.5 eV.
The observed doping can be explained from 
geometrical arguments. The bounded oxygen and titanium
atoms can induce p- or n-type doping when close to
the graphene surface. The TiO$_{2}$ monolayer in the g2x2-hex-TiO$_{2}$ structure
can be seen as consisting of three layers --
a titanium layer sandwiched between two oxygen layers,
thus exposing oxygen atoms to graphene.
Similar charge transfers have been observed at the interface
between graphene and rutile TiO$_{2}$ (110), where
hole doping in graphene and an accompanying shift
of the Fermi level relative to the Dirac cone was observed.\ \cite{du_2011-g_on_tio2}

The aforementioned reasoning proves to be valid also
for the other systems we have investigated. The model structures
in Figs. \ref{fig_tio2_g_structures}(a) and
\ref{fig_tio2_g_structures}(b) differ by the amount of
oxygen or titanium atoms they expose to the graphene
surface. The charge population analysis as well as the
electronic band structure provide consistent evidence
of p- or n-type doping depending on whether
oxygen or titanium atoms are closer to the carbon atoms, respectively.
The shift of the Dirac cone with respect to the Fermi level
is -1.69 eV for g2x2-aligned hex-TiO$_{2}$ 
and 0 eV for g-sq8-TiO$_{2}$ structure, and
the amount of charge, calculated with iterative Hirshfeld
method, transfered from graphene
is -0.452 e and 0.026 e, respectively.

Recently it was found experimentally that
the graphene gate dependent conductivity
can be recovered after oxidation of adsorbed Ti atoms. \cite{mccreary_2011}
This recovery can be explained with
the autocompensation of the titanium dioxide surface\ \cite{diebold_2003}
exposed to graphene. We showed above that
TiO$_{2}$ structures can induce both p- or n-type
doping in graphene thus opening the possibility
of mutual compensation of the doping effects.
Although the groundstate TiO$_{2}$ monolayer induces
hole doping to graphene the performed
calculations suggest that metastable states
of TiO$_{2}$ monolayers can be responsible for
the recovery of the graphene gate dependent conductivity.
However, the geometrical nature of the autocompensation
does not imply the actual presence
of the above considered monolayer TiO$_{2}$ crystals 
on the graphene surface in the experiment.
Existence of any TiO$_{2}$ nanostructure with both
p- or n-type doping sides exposed to the graphene
surface will lead to similar effects.

\section{Conclusions}

We have demonstrated strong n-type doping
of Ti covered graphene. While the various charge population
analysis methods failed to provide reasonable quantitative
results for the shift of the Dirac point with respect to the Fermi level
our investigation of the increased
difference of the work functions ($\Phi_{M} - \Phi_{G}$)
provided a consistent picture of the character
of the induced doping. Titanium is found to be strongly chemisorbed
on graphene and the highest monolayer
coverage is found to have stoichiometry Ti$_{3}$C$_{8}$ .
The studied TiO$_{2}$ monolayer crystals aligned on
top of graphene induce charge doping that depends on the nature of the
atoms that are closest to graphene with oxygen acting as an acceptor
and titanium as a donor. The ground state structure 
of TiO$_{2}$ monolayer covered graphene exhibits p-type
doping of graphene, nevertheless the calculations suggest
these metastable states of TiO$_{2}$ monolayer can be 
responsible for the recovery of the graphene
gate dependent conductivity after oxidation
of adsorbed titanium, as recently found in experiment.\ \cite{mccreary_2011}.

\begin{acknowledgments}
This work is supported by the ESF-Eurocores program EuroGRAPHENE (project CONERAN) and the Flemish Science Foundation (FWO-Vl). This work was carried out using the HPC infrastructure of the University of Antwerp (CalcUA) a division of the Flemish Supercomputer Center (VSC), which is funded by the Hercules foundation.
\end{acknowledgments}


\begin{thebibliography}{22}
\expandafter\ifx\csname natexlab\endcsname\relax\def\natexlab#1{#1}\fi
\expandafter\ifx\csname bibnamefont\endcsname\relax
  \def\bibnamefont#1{#1}\fi
\expandafter\ifx\csname bibfnamefont\endcsname\relax
  \def\bibfnamefont#1{#1}\fi
\expandafter\ifx\csname citenamefont\endcsname\relax
  \def\citenamefont#1{#1}\fi
\expandafter\ifx\csname url\endcsname\relax
  \def\url#1{\texttt{#1}}\fi
\expandafter\ifx\csname urlprefix\endcsname\relax\def\urlprefix{URL }\fi
\providecommand{\bibinfo}[2]{#2}
\providecommand{\eprint}[2][]{\url{#2}}

\bibitem[{\citenamefont{Novoselov et~al.}(2004)\citenamefont{Novoselov, Geim,
  Morozov, Jiang, Zhang, Dubonos, Grigorieva, and Firsov}}]{novoselov_2004}
\bibinfo{author}{\bibfnamefont{K.~S.} \bibnamefont{Novoselov}},
  \bibinfo{author}{\bibfnamefont{A.~K.} \bibnamefont{Geim}},
  \bibinfo{author}{\bibfnamefont{S.~V.} \bibnamefont{Morozov}},
  \bibinfo{author}{\bibfnamefont{D.}~\bibnamefont{Jiang}},
  \bibinfo{author}{\bibfnamefont{Y.}~\bibnamefont{Zhang}},
  \bibinfo{author}{\bibfnamefont{S.~V.} \bibnamefont{Dubonos}},
  \bibinfo{author}{\bibfnamefont{I.~V.} \bibnamefont{Grigorieva}},
  \bibnamefont{and} \bibinfo{author}{\bibfnamefont{A.~A.}
  \bibnamefont{Firsov}}, \bibinfo{journal}{Science}
  \textbf{\bibinfo{volume}{306}}, \bibinfo{pages}{666} (\bibinfo{year}{2004}),
  \eprint{http://www.sciencemag.org/content/306/5696/666.full.pdf},
  \urlprefix\url{http://www.sciencemag.org/content/306/5696/666.abstract}.

\bibitem[{\citenamefont{Robinson et~al.}(2011)\citenamefont{Robinson, LaBella,
  Zhu, Hollander, Kasarda, Hughes, Trumbull, Cavalero, and
  Snyder}}]{robinson_2011}
\bibinfo{author}{\bibfnamefont{J.~A.} \bibnamefont{Robinson}},
  \bibinfo{author}{\bibfnamefont{M.}~\bibnamefont{LaBella}},
  \bibinfo{author}{\bibfnamefont{M.}~\bibnamefont{Zhu}},
  \bibinfo{author}{\bibfnamefont{M.}~\bibnamefont{Hollander}},
  \bibinfo{author}{\bibfnamefont{R.}~\bibnamefont{Kasarda}},
  \bibinfo{author}{\bibfnamefont{Z.}~\bibnamefont{Hughes}},
  \bibinfo{author}{\bibfnamefont{K.}~\bibnamefont{Trumbull}},
  \bibinfo{author}{\bibfnamefont{R.}~\bibnamefont{Cavalero}}, \bibnamefont{and}
  \bibinfo{author}{\bibfnamefont{D.}~\bibnamefont{Snyder}},
  \bibinfo{journal}{Appl. Phys. Lett.} \textbf{\bibinfo{volume}{98}},
  \bibinfo{eid}{053103} (pages~\bibinfo{numpages}{3}) (\bibinfo{year}{2011}),
  \urlprefix\url{http://link.aip.org/link/?APL/98/053103/1}.

\bibitem[{\citenamefont{Dag et~al.}(2004)\citenamefont{Dag, Durgun, and
  Ciraci}}]{dag_2004}
\bibinfo{author}{\bibfnamefont{S.}~\bibnamefont{Dag}},
  \bibinfo{author}{\bibfnamefont{E.}~\bibnamefont{Durgun}}, \bibnamefont{and}
  \bibinfo{author}{\bibfnamefont{S.}~\bibnamefont{Ciraci}},
  \bibinfo{journal}{Phys. Rev. B} \textbf{\bibinfo{volume}{69}},
  \bibinfo{pages}{121407} (\bibinfo{year}{2004}),
  \urlprefix\url{http://link.aps.org/doi/10.1103/PhysRevB.69.121407}.

\bibitem[{\citenamefont{Carrillo et~al.}(2009)\citenamefont{Carrillo, Rangel,
  and Maga{\~{n}}a}}]{carrillo_2009}
\bibinfo{author}{\bibfnamefont{I.}~\bibnamefont{Carrillo}},
  \bibinfo{author}{\bibfnamefont{E.}~\bibnamefont{Rangel}}, \bibnamefont{and}
  \bibinfo{author}{\bibfnamefont{L.~F.} \bibnamefont{Maga{\~{n}}a}},
  \bibinfo{journal}{Carbon} \textbf{\bibinfo{volume}{47}},
  \bibinfo{pages}{2758} (\bibinfo{year}{2009}), ISSN \bibinfo{issn}{0008-6223},
  \urlprefix\url{http://www.sciencedirect.com/science/article/pii/S0008622309003728}.

\bibitem[{\citenamefont{Yildirim and Ciraci}(2005)}]{yildirim_2005}
\bibinfo{author}{\bibfnamefont{T.}~\bibnamefont{Yildirim}} \bibnamefont{and}
  \bibinfo{author}{\bibfnamefont{S.}~\bibnamefont{Ciraci}},
  \bibinfo{journal}{Phys. Rev. Lett.} \textbf{\bibinfo{volume}{94}},
  \bibinfo{pages}{175501} (\bibinfo{year}{2005}),
  \urlprefix\url{http://link.aps.org/doi/10.1103/PhysRevLett.94.175501}.

\bibitem[{\citenamefont{Liu et~al.}(2010)\citenamefont{Liu, Ren, He, and
  Cheng}}]{liu_2010}
\bibinfo{author}{\bibfnamefont{Y.}~\bibnamefont{Liu}},
  \bibinfo{author}{\bibfnamefont{L.}~\bibnamefont{Ren}},
  \bibinfo{author}{\bibfnamefont{Y.}~\bibnamefont{He}}, \bibnamefont{and}
  \bibinfo{author}{\bibfnamefont{H.-P.} \bibnamefont{Cheng}},
  \bibinfo{journal}{J. Phys.: Condens. Matter} \textbf{\bibinfo{volume}{22}},
  \bibinfo{pages}{445301} (\bibinfo{year}{2010}),
  \urlprefix\url{http://stacks.iop.org/0953-8984/22/i=44/a=445301}.

\bibitem[{\citenamefont{Rojas and Leiva}(2007)}]{rojas_2007}
\bibinfo{author}{\bibfnamefont{M.~I.} \bibnamefont{Rojas}} \bibnamefont{and}
  \bibinfo{author}{\bibfnamefont{E.~P.~M.} \bibnamefont{Leiva}},
  \bibinfo{journal}{Phys. Rev. B} \textbf{\bibinfo{volume}{76}},
  \bibinfo{pages}{155415} (\bibinfo{year}{2007}),
  \urlprefix\url{http://link.aps.org/doi/10.1103/PhysRevB.76.155415}.

\bibitem[{\citenamefont{McCreary et~al.}(2011)\citenamefont{McCreary, Pi, and
  Kawakami}}]{mccreary_2011}
\bibinfo{author}{\bibfnamefont{K.~M.} \bibnamefont{McCreary}},
  \bibinfo{author}{\bibfnamefont{K.}~\bibnamefont{Pi}}, \bibnamefont{and}
  \bibinfo{author}{\bibfnamefont{R.~K.} \bibnamefont{Kawakami}},
  \bibinfo{journal}{Appl. Phys. Lett.} \textbf{\bibinfo{volume}{98}},
  \bibinfo{eid}{192101} (\bibinfo{year}{2011}),
  \urlprefix\url{http://link.aip.org/link/?APL/98/192101/1}.

\bibitem[{\citenamefont{Bl{\"{o}}chl}(1994)}]{bloch_1994-paw}
\bibinfo{author}{\bibfnamefont{P.~E.} \bibnamefont{Bl{\"{o}}chl}},
  \bibinfo{journal}{Phys. Rev. B} \textbf{\bibinfo{volume}{50}},
  \bibinfo{pages}{17953} (\bibinfo{year}{1994}),
  \urlprefix\url{http://link.aps.org/doi/10.1103/PhysRevB.50.17953}.

\bibitem[{\citenamefont{Monkhorst and Pack}(1976)}]{monkhorst_1976}
\bibinfo{author}{\bibfnamefont{H.~J.} \bibnamefont{Monkhorst}}
  \bibnamefont{and} \bibinfo{author}{\bibfnamefont{J.~D.} \bibnamefont{Pack}},
  \bibinfo{journal}{Phys. Rev. B} \textbf{\bibinfo{volume}{13}},
  \bibinfo{pages}{5188} (\bibinfo{year}{1976}),
  \urlprefix\url{http://link.aps.org/doi/10.1103/PhysRevB.13.5188}.

\bibitem[{\citenamefont{Bultinck et~al.}(2007)\citenamefont{Bultinck, Alsenoy,
  Ayers, and Carb{\'{o}}-Dorca}}]{bultinck_2007-ih}
\bibinfo{author}{\bibfnamefont{P.}~\bibnamefont{Bultinck}},
  \bibinfo{author}{\bibfnamefont{C.~V.} \bibnamefont{Alsenoy}},
  \bibinfo{author}{\bibfnamefont{P.~W.} \bibnamefont{Ayers}}, \bibnamefont{and}
  \bibinfo{author}{\bibfnamefont{R.}~\bibnamefont{Carb{\'{o}}-Dorca}},
  \bibinfo{journal}{J. Chem. Phys.} \textbf{\bibinfo{volume}{126}},
  \bibinfo{eid}{144111} (\bibinfo{year}{2007}),
  \urlprefix\url{http://link.aip.org/link/?JCP/126/144111/1}.

\bibitem[{\citenamefont{Sevin{\c{c}}li
  et~al.}(2008)\citenamefont{Sevin{\c{c}}li, Topsakal, Durgun, and
  Ciraci}}]{sevincli_2008-tm_on_g}
\bibinfo{author}{\bibfnamefont{H.}~\bibnamefont{Sevin{\c{c}}li}},
  \bibinfo{author}{\bibfnamefont{M.}~\bibnamefont{Topsakal}},
  \bibinfo{author}{\bibfnamefont{E.}~\bibnamefont{Durgun}}, \bibnamefont{and}
  \bibinfo{author}{\bibfnamefont{S.}~\bibnamefont{Ciraci}},
  \bibinfo{journal}{Phys. Rev. B} \textbf{\bibinfo{volume}{77}},
  \bibinfo{pages}{195434} (\bibinfo{year}{2008}),
  \urlprefix\url{http://link.aps.org/doi/10.1103/PhysRevB.77.195434}.

\bibitem[{\citenamefont{Rangel et~al.}(2009)\citenamefont{Rangel,
  Ruiz-Chavarria, and Magana}}]{rangel_2009-h2o_on_ti-g}
\bibinfo{author}{\bibfnamefont{E.}~\bibnamefont{Rangel}},
  \bibinfo{author}{\bibfnamefont{G.}~\bibnamefont{Ruiz-Chavarria}},
  \bibnamefont{and} \bibinfo{author}{\bibfnamefont{L.~F.}
  \bibnamefont{Magana}}, \bibinfo{journal}{Carbon}
  \textbf{\bibinfo{volume}{47}}, \bibinfo{pages}{531} (\bibinfo{year}{2009}),
  ISSN \bibinfo{issn}{0008-6223},
  \urlprefix\url{http://www.sciencedirect.com/science/article/pii/S0008622308006519}.

\bibitem[{\citenamefont{Zanella et~al.}(2008)\citenamefont{Zanella, Fagan,
  Mota, and Fazzio}}]{zanella_2008-elmag_of_ti_fe_on_g}
\bibinfo{author}{\bibfnamefont{I.}~\bibnamefont{Zanella}},
  \bibinfo{author}{\bibfnamefont{S.~B.} \bibnamefont{Fagan}},
  \bibinfo{author}{\bibfnamefont{R.}~\bibnamefont{Mota}}, \bibnamefont{and}
  \bibinfo{author}{\bibfnamefont{A.}~\bibnamefont{Fazzio}},
  \bibinfo{journal}{J. Phys. Chem. C} \textbf{\bibinfo{volume}{112}},
  \bibinfo{pages}{9163} (\bibinfo{year}{2008}),
  \eprint{http://pubs.acs.org/doi/pdf/10.1021/jp711691r},
  \urlprefix\url{http://pubs.acs.org/doi/abs/10.1021/jp711691r}.

\bibitem[{\citenamefont{Leenaerts et~al.}(2008)\citenamefont{Leenaerts,
  Partoens, and Peeters}}]{leenaerts_2008-charge}
\bibinfo{author}{\bibfnamefont{O.}~\bibnamefont{Leenaerts}},
  \bibinfo{author}{\bibfnamefont{B.}~\bibnamefont{Partoens}}, \bibnamefont{and}
  \bibinfo{author}{\bibfnamefont{F.~M.} \bibnamefont{Peeters}},
  \bibinfo{journal}{Appl. Phys. Lett.} \textbf{\bibinfo{volume}{92}}
  (\bibinfo{year}{2008}), ISSN \bibinfo{issn}{0003-6951}.

\bibitem[{\citenamefont{Bader}(1991)}]{bader_1991}
\bibinfo{author}{\bibfnamefont{R.~F.~W.} \bibnamefont{Bader}},
  \bibinfo{journal}{Chem. Rev.} \textbf{\bibinfo{volume}{91}},
  \bibinfo{pages}{893} (\bibinfo{year}{1991}),
  \eprint{http://pubs.acs.org/doi/pdf/10.1021/cr00005a013},
  \urlprefix\url{http://pubs.acs.org/doi/abs/10.1021/cr00005a013}.

\bibitem[{\citenamefont{Tang et~al.}(2009)\citenamefont{Tang, Sanville, and
  Henkelman}}]{tang_2009}
\bibinfo{author}{\bibfnamefont{W.}~\bibnamefont{Tang}},
  \bibinfo{author}{\bibfnamefont{E.}~\bibnamefont{Sanville}}, \bibnamefont{and}
  \bibinfo{author}{\bibfnamefont{G.}~\bibnamefont{Henkelman}},
  \bibinfo{journal}{J. Phys.: Condens. Matter} \textbf{\bibinfo{volume}{21}},
  \bibinfo{pages}{084204} (\bibinfo{year}{2009}),
  \urlprefix\url{http://stacks.iop.org/0953-8984/21/i=8/a=084204}.

\bibitem[{\citenamefont{Fonseca~Guerra
  et~al.}(2004)\citenamefont{Fonseca~Guerra, Handgraaf, Baerends, and
  Bickelhaupt}}]{celia_2004-voronoi}
\bibinfo{author}{\bibfnamefont{C.}~\bibnamefont{Fonseca~Guerra}},
  \bibinfo{author}{\bibfnamefont{J.-W.} \bibnamefont{Handgraaf}},
  \bibinfo{author}{\bibfnamefont{E.~J.} \bibnamefont{Baerends}},
  \bibnamefont{and} \bibinfo{author}{\bibfnamefont{F.~M.}
  \bibnamefont{Bickelhaupt}}, \bibinfo{journal}{J. Comput. Chem.}
  \textbf{\bibinfo{volume}{25}}, \bibinfo{pages}{189} (\bibinfo{year}{2004}),
  ISSN \bibinfo{issn}{1096-987X}.

\bibitem[{\citenamefont{Giovannetti et~al.}(2008)\citenamefont{Giovannetti,
  Khomyakov, Brocks, Karpan, van~den Brink, and Kelly}}]{giovannetti_2008}
\bibinfo{author}{\bibfnamefont{G.}~\bibnamefont{Giovannetti}},
  \bibinfo{author}{\bibfnamefont{P.~A.} \bibnamefont{Khomyakov}},
  \bibinfo{author}{\bibfnamefont{G.}~\bibnamefont{Brocks}},
  \bibinfo{author}{\bibfnamefont{V.~M.} \bibnamefont{Karpan}},
  \bibinfo{author}{\bibfnamefont{J.}~\bibnamefont{van~den Brink}},
  \bibnamefont{and} \bibinfo{author}{\bibfnamefont{P.~J.} \bibnamefont{Kelly}},
  \bibinfo{journal}{Phys. Rev. Lett.} \textbf{\bibinfo{volume}{101}},
  \bibinfo{pages}{026803} (\bibinfo{year}{2008}),
  \urlprefix\url{http://link.aps.org/doi/10.1103/PhysRevLett.101.026803}.

\bibitem[{\citenamefont{Evarestov et~al.}(2010)\citenamefont{Evarestov,
  Bandura, and Losev}}]{evarestov_2010}
\bibinfo{author}{\bibfnamefont{R.}~\bibnamefont{Evarestov}},
  \bibinfo{author}{\bibfnamefont{A.}~\bibnamefont{Bandura}}, \bibnamefont{and}
  \bibinfo{author}{\bibfnamefont{M.}~\bibnamefont{Losev}},
  \bibinfo{journal}{Russ. J. Gen. Chem.} \textbf{\bibinfo{volume}{80}},
  \bibinfo{pages}{1152} (\bibinfo{year}{2010}), ISSN \bibinfo{issn}{1070-3632},
  \bibinfo{note}{10.1134/S1070363210060198},
  \urlprefix\url{http://dx.doi.org/10.1134/S1070363210060198}.

\bibitem[{\citenamefont{Du et~al.}(2011)\citenamefont{Du, Ng, Bell, Zhu, Amal,
  and Smith}}]{du_2011-g_on_tio2}
\bibinfo{author}{\bibfnamefont{A.}~\bibnamefont{Du}},
  \bibinfo{author}{\bibfnamefont{Y.~H.} \bibnamefont{Ng}},
  \bibinfo{author}{\bibfnamefont{N.~J.} \bibnamefont{Bell}},
  \bibinfo{author}{\bibfnamefont{Z.}~\bibnamefont{Zhu}},
  \bibinfo{author}{\bibfnamefont{R.}~\bibnamefont{Amal}}, \bibnamefont{and}
  \bibinfo{author}{\bibfnamefont{S.~C.} \bibnamefont{Smith}},
  \bibinfo{journal}{J. Phys. Chem. Lett.} \textbf{\bibinfo{volume}{2}},
  \bibinfo{pages}{894} (\bibinfo{year}{2011}),
  \eprint{http://pubs.acs.org/doi/pdf/10.1021/jz2002698},
  \urlprefix\url{http://pubs.acs.org/doi/abs/10.1021/jz2002698}.

\bibitem[{\citenamefont{Diebold}(2003)}]{diebold_2003}
\bibinfo{author}{\bibfnamefont{U.}~\bibnamefont{Diebold}},
  \bibinfo{journal}{Surf. Sci. Rep.} \textbf{\bibinfo{volume}{48}},
  \bibinfo{pages}{53} (\bibinfo{year}{2003}), ISSN \bibinfo{issn}{0167-5729},
  \urlprefix\url{http://www.sciencedirect.com/science/article/pii/S0167572902001000}.

\end{thebibliography}
\end{document}